\documentclass[preprint,12pt]{elsarticle}

\usepackage{graphicx}
\usepackage{amssymb}
\usepackage{amsmath}

\newcommand{\up}{\partial}
\newcommand{\uc}{\overline}
\newcommand{\ud}{\,{\mathrm d}}
\newcommand{\zc}{\overline{z}}
\newcommand{\wc}{\overline{w}}

\newcommand{\uRe}{{\operatorname{Re}}}

\newcommand{\Res}{\operatorname {Res}}

\journal{Journal of Magnetism and Magnetic Materials}

\begin{document}

\begin{frontmatter}

\title{Analytical approximations to the core radius and energy of magnetic
vortex in thin ferromagnetic disks}

\author{Konstantin L. Metlov}

%\email{metlov@fti.dn.ua}
\ead{metlov@fti.dn.ua}

%\affiliation{Donetsk Institute for Physics and Technology NAS, Donetsk, Ukraine 83114}
\address{Donetsk Institute for Physics and Technology NAS, Donetsk, Ukraine 83114}
\date{\today}
\begin{abstract}
The energy of magnetic vortex core and its equilibrium radius in thin
circular cylinder were first presented by Usov and Peschany in 1994.
Yet, the magnetostatic function, entering the energy expression, is
hard to evaluate and approximate. Here, precise and explicit
analytical approximations to this function (as well as equilibrium
vortex core radius and energy) are derived in terms of elementary
functions. Also, several simplifying approximations to the magnetic
Hamiltonian and their impact on theoretical stability of magnetic
vortex state are discussed.
\end{abstract}
%\pacs{75.60.Ch, 75.70.Kw, 85.70.Kh}
%\keywords{micromagnetics, magnetic nano-dots, magnetic vortex}
\begin{keyword}
micromagnetics \sep magnetic nano-dots \sep magnetic vortex
\end{keyword}
\end{frontmatter}

%\maketitle
\section{Introduction }
The first topological soliton was discovered as a solution of
non-linear field theory equations by Skyrme\cite{Skyrme62}. It had a
form of three-dimensional hedgehog and was named subsequently
``skyrmion'' in honor of the discoverer. After the landmark work of
Belavin and Polyakov\cite{BP75}, topological solitons have crossed the
boundary into condensed matter physics. The latter authors discovered
much more topological soliton solutions in the infinite Heisenberg
ferromagnet, as many as there are rational functions of complex
variable, mapping any such function into some equilibrium magnetic
structure. Zeros of numerator (possibly with higher multiplicity) of
these rational functions correspond to the centers of magnetic
vortices, while zeros of denominator to the centers of magnetic
anti-vortices. In the model of infinite 2D ferromagnet, considered by
Belavin and Polyakov, solitons are absolutely stable. Once magnetic
texture with a certain topological charge (number of magnetic
vortices) is created, all the structures with different topological
charge are separated by an infinite energy barrier. It is worth noting
that 7 years earlier essentially the same mathematics of rational
functions of complex variable was applied to the problem of magnetic
singularities (Bloch points) in 3D ferromagnet by
D\"oring\cite{doring68}, who also found that the exchange energy of
ferromagnet around a Bloch point depends only on degrees of numerator
and denominator of the corresponding rational function. His energy
expression is exactly the same as that of Belavin and
Polyakov\cite{BP75} for 2D ferromagnet. However (probably, because
Bloch points, to which model of Ref.~\cite{doring68} applies, are
rather exotic objects in magnetism), the paper by D\"oring is
currently much less known and cited.

The model of Belavin and Polyakov became known in particle physics as
non-linear O(3) $\sigma$ model in 3+1 dimensions and reformulated
elegantly in terms of functions of complex variable by
G. Woo\cite{Woo77}. David J. Gross found additional family of
``meron'' solutions to it\cite{G78}. Since then, the original
Belavin-Polyakov solutions became known as just ``solitons''. Merons,
and all other O(3) $\sigma$ model solutions besides
solitons\cite{Woo77}, have infinite energy in unbounded 2-d
ferromagnet, but can be realized when the ferromagnet is
finite\cite{M01_solitons}.

While these solutions were obtained long ago, the question of their
stability has a history of its own. Kosterlitz and Thouless\cite{KT73}
analyzed stability of planar vortices in 2D ferromagnet and came to
conclusion that they are unstable and such order could not exist. It
is, indeed, true that the energy of Belavin-Polyakov solitons is
scale-invariant and their size is, thus, undefined. In real
ferromagnets, however, there are various other interactions (not
exotic at all), which make the vortices stable. Usov and Peschany,
were first to show that dipolar magnetostatic interaction stabilizes
magnetic vortex in ferromagnetic cylinder both with respect to core
radius change\cite{UP93} and vortex center
displacement\cite{UP94}. Their results were later fully confirmed
experimentally, starting with the direct observation of magnetic
vortex core and measurement of its radius\cite{SOHSO00}. These and
following experiments made single vortex state not only interesting
from fundamental point of view, but also an essential component of
emerging spintronic devices (such as MRAM elements, based on vortex
core polarity\cite{VanWaeyenberge2006} or chirality\cite{AO2009}
switching, or spin-polarized current magnetic
nano-oscillators\cite{Pribiag_Vortex_NatPhys07}). It is also
prerequisite for study of more complex multi-vortex magnetic
configurations in planar nano-elements of various shapes.

Here, starting from recent (and more general) description of
magnetization distributions in finite nano-elements via functions of
complex variable \cite{M10} the impact of various approximations on
vortex stability is reviewed in unified manner and the expression for
vortex core radius in circular cylinder\cite{UP93} is re-derived. It
defines the core radius implicitly via an equation and an integral of
certain special functions, which is very inconvenient to evaluate and
approximate at small cylinder thickness because in this limit it is
not analytic and its higher derivatives do not exist. It is, however,
possible to introduce small parameters and expand the special
functions and the vortex core radius into series, obtaining an
explicit analytical approximate (but very precise) expressions,
presented at the end.

\section{Magnetic vortex in complex variables and its exchange energy}
In finite planar nano-elements the equilibrium magnetization
configurations can be described via rational functions of complex
variable with real coefficients\cite{M10} (as opposed to complex
coefficients in the case of infinite film\cite{BP75,Woo77}). The
simplest ansatz for magnetic vortex in circular cylinder (of thickness
$L_Z$ and radius $R$) can be written in the complex notation as
\begin{equation}
 \label{eq:f}
 f(z)=\imath (z-a)/R_V
\end{equation}
where $z= X + \imath Y$ with $X$ and $Y$ being the Cartesian coordinates in
the cylinder's plane (the magnetization distribution is assumed to be
independent
on out-of-plane coordinate $Z$), $R_V$ is the vortex core radius and $a$ is the
displacement of the vortex from the origin ($a=0$ corresponds to the centered
vortex). Let us then define a complex function
\begin{equation}
  \label{eq:join}
 w(z, \zc)=\left\{
  \begin{array}{ll}
  f(z) & |f(z)| \leq 1 \\
  f(z)/|f(z)| & |f(z)|>1
  \end{array}
\right. ,
\end{equation}
where the line over variable denotes complex conjugation. The function $w$ is
shown to depend explicitly on both $z$ and $\zc$ because it is, in general, not
holomorphic. It consists of two parts: soliton (where it is analytic and $\up
w(z, \zc)/\up \zc=0$) and meron (where $w \wc = 1$, joined at a line (possibly
multiply-connected if there are several vortices or anti-vortices) $|f|=1$. The
magnetization components, normalized by material's saturation magnetization
$M_S$, are then expressed via stereographic projection as
\begin{eqnarray}
  \label{eq:mxmy}
 m_x + \imath m_y & = & \frac{2 w(z,\zc)}{1+w(z,\zc)\wc(z,\zc)} \\
  \label{eq:mz}
 m_z & = & \frac{1-w(z,\zc)\wc(z,\zc)}{1+w(z,\zc)\wc(z,\zc)}.
\end{eqnarray}
Being written via the magnetization components in
Eq. (\ref{eq:mxmy})-(\ref{eq:mz}), the ansatz in Eq.(\ref{eq:f}) is
exactly equivalent to the one by Usov and Peschany\cite{UP93} and also
belongs to the class of trial functions, considered by Kosterlitz and
Thouless\cite{KT73}. Following the latter work, let us first take into
account only the exchange interaction. In complex notation the
exchange energy density (omitting the factor $C/2$, where $C$ is the
exchange stiffness) can be directly expressed via the function $w$:
\begin{equation}
  \label{eq:exchange}
  \sum_{i=x,y,z} (\vec{\nabla} m_i)^2 = 
  \frac{8}{(1+w \uc w)^2}
  \left(
    \frac{\up w}{\up z} \frac{\up \uc w}{\up \uc z}+
    \frac{\up w}{\up \uc z} \frac{\up \uc w}{\up z}
  \right),
\end{equation}
where $\partial/\partial z = (\partial /\partial X - \imath\,\partial
/\partial Y)/2$ and $\partial/\partial \zc = (\partial /\partial X + \imath\,
\partial /\partial Y)/2$. The total exchange energy can be obtained by
integrating the density (\ref{eq:exchange}) over
nano-element's volume.
Recalling the Riemann-Greene theorem
\begin{equation}
  \label{eq:Riemann-Greene}
  \frac{1}{2\imath}\oint_{\up \cal D} u(\zeta,\overline{\zeta}) 
  \ud \zeta =
  \iint_{\cal D} \frac{\partial u(z,\overline{z})}{\partial \overline{z}}
  \ud X \ud Y,
\end{equation}
where $u$ is a complex function of the complex argument (not necessary
analytic\footnote{For analytic $u$ the double integral over ${\cal D}$
  is equal to $0$, which is the manifestation of Cauchy theorem.}), it
is possible to reduce the area integral over cylinder's face $\cal D$
for the total exchange energy to a contour integral over its boundary
$\up \cal D$, provided there is a complex function, whose derivative
over $\zc$ yields the exchange energy density (\ref{eq:exchange}).
Luckily, such function (actually two functions, one for soliton and
one for meron part of $w$) can be easily obtained by direct
integration of (\ref{eq:exchange}) with $w$ from each of the
conditions in (\ref{eq:join}):
\begin{eqnarray}
  \label{eq:exchange_uz_soliton}
  u^S(z,\overline z) &=& - \frac{8}{1+f(z)\overline{f}(\overline{z})}
  \frac{1}{f(z)} \frac{\partial f}{\partial z}, \\
  \label{eq:exchange_uz_meron}
  u^{\mathrm M}(z,\overline z) &=&
  \frac{1}{f(z)} \frac{\partial f}{\partial z} 
  \log \left( f(z) \overline{f}(\overline{z}) \right).
\end{eqnarray}
Thus, from
(\ref{eq:Riemann-Greene}), the total 
exchange energy inside the soliton
is
\begin{equation}
  \label{eq:exchange_soliton_general}
  \frac{E_\mathrm{EX}^{\mathrm S}}{C L_Z/2}=
  \frac{2}{\imath}\oint_{|f(\zeta)|=1}\frac{1}{f(\zeta)}
  \frac{\partial f(\zeta)}{\partial \zeta} \ud \zeta,
\end{equation}
where the fact that $|f(\zeta)|=1$ on the integration contour is used
and the additional minus sign appears because the original
contour of integration has to be walked clockwise. The function under
the integral is analytic everywhere except the vortex centers $z_i$, where
$f(z_i)=0$. Assuming that line $|f(\zeta)|=1$ does not cross the particle
boundary, it is possible to tighten the contours around each topological
singularity (vortex or anti-vortex center) and use the residue theorem
\begin{equation}
  \label{eq:exchange_soliton_final}
  \frac{E_\mathrm{EX}^{\mathrm S}}{C L_Z/2}= 4 \pi \sum_{i} 
  \Res
  \left.
    \frac{1}{f(z)}\frac{\partial f(z)}{\partial z}
  \right|_{z\rightarrow z_i}.
\end{equation}
In particular, for $f(z)$ from Eq.~\ref{eq:f} this gives $E^{\mathrm
  S}/(C L_Z/2)= 4\pi$. If there are several vortices inside the
particle, the energy will be multiplied by their total number,
including multiplicities.

For the meron part, the integration boundary is multiply-connected.
However, on the inner boundaries (encircling solitons) $|f(\zeta)|=1$ and
 $u^{\mathrm M}(z,\overline z) \sim \log 1 =0$. Thus, only the integral over the
cylinder's outer boundary remains
\begin{equation}
  \label{eq:exchange_meron_general}
  \frac{E_\mathrm{EX}^{\mathrm M}}{C L_Z/2}= 
  \frac{1}{2\imath} \oint_{\up \cal D}
  \frac{1}{f(\zeta)}\frac{\partial f(\zeta)}{\partial \zeta}
  \log f(\zeta)\overline{f}(\overline{\zeta}) \ud \zeta
\end{equation}
for $f(z)$ from Eq.~\ref{eq:f} and the nano-element, shaped as circular
cylinder ($\partial \cal D$ is $|z|=R$)
\begin{eqnarray}
  \label{eq:exchange_meron_uniform}
  \frac{E_\mathrm{EX}^{\mathrm M}}{C L_Z/2} & = &
  \!\!\!\int\limits_0^{2\pi} \frac{(1\!-\!a \cos (\varphi)/R) \log \!\left(\frac{a^2-2
a R \cos (\varphi)+R^2}{R_V^2}\right)}{2 \left(a^2/R^2-2 a \cos
   (\varphi)/R+1\right)} \ud \varphi \nonumber \\
  & = & 
  \pi  \log \left(1-\frac{a^2}{R^2}\right)-2 \pi  \log
   \left(\frac{R_V}{R}\right),
\end{eqnarray}
and the total exchange energy 
$e_{\mathrm{EX}}=(E^{\mathrm S}+E^{\mathrm M})/(\mu_0\gamma_B M_S^2 \pi R^2 L_Z)$
(in subsequent text all the dimensionless energies, denoted by small letter $e$ 
with different sub-/superscripts use the same normalization) is
\begin{equation}
  \label{eq:exchange_total}
  e_{\mathrm{EX}}=\frac{L_E^2}{R^2}\left(2-\log \frac
{R_V}{R}+\log\sqrt{1-\frac{a^2}{R^2}}\right),
\end{equation}
where $\gamma_B=4\pi$, $\mu_0=1$ in CGS units and $\gamma_B=1$ in
SI\cite{AharoniBook} and the exchange length\footnote{The other
  common definition of the exchange length
  $L_E^\mathrm{UP}=\sqrt{C/(\mu_0 M_S^2)}$, used by Usov and
  Peschany\cite{UP93} and in many followup works, is, actually,
  dependent on system of measurement units and makes the formulas for
  the dimensionless energy and all the derived quantities depend on
  units too. To avoid this complication the definition
  $L_E=\sqrt{C/(\mu_0\gamma_B M_S^2)}$ is adopted here, which in CGS
  units (which are almost exclusively used in conjunction with
  $L_E^\mathrm{UP}$ definition) is by factor $\sqrt{4\pi}$ smaller
  (making all the lengths, measured in units of $L_E$, by factor
  $\sqrt{4\pi}$ larger then the lengths, measured in
  $L_E^\mathrm{UP}$). The numeric quantities here are given for both
  definitions of the exchange length to make the comparison to other
  results in the literature easier.} $L_E=\sqrt{C/(\mu_0
  \gamma_B M_S^2)}$.  It can be seen immediately that
the exchange energy decreases with increasing of the vortex core size
$R_V$. The expression (\ref{eq:exchange_total}) is, formally, valid
only for $R_V<R$, but it can be easily shown that the energy continues
to decrease for larger $R_V$, reaching equilibrium for
$R_V\rightarrow\infty$. This confirms the conclusion of Kosterlitz and
Thouless\cite{KT73} that magnetic vortices are unstable when only the
exchange interaction is taken into account.

\section{Magnetostatic energy of magnetic vortex}
The long-range dipolar interaction between the local magnetic moments
is present in all magnets. Strictly speaking, it is not instantaneous
and its speed is limited by the speed of light. The account for
retardation effects, however, contributes to the
dissipation\cite{BT11}. It is convenient (if magnetic nano-elements
are small enough and characteristic timescales are large enough) to
consider the dipolar interaction in magnetostatic approximation,
making it non-local. Non-locality still poses a major mathematical
difficulty, since it makes the equations for equilibrium magnetization
distribution not only non-linear partial differential, but also
integral.  To alleviate this difficulty a number of local
magnetostatic approximations had been developed. The most common (and
very useful for considering domain walls in thin films) is based on
using the local uniaxial in-plane anisotropy term $K_\mathrm{MS}
m_Z^2$ instead of magnetostatic interaction. Selecting the local bulk
anisotropy $K_\mathrm{MS}=\mu_0 \gamma_B M_S^2/2$, one gets the exact
correspondence between the approximate and exact magnetostatic energy
density in two limiting cases: when the film is magnetized in-plane
(magnetostatic energy is $0$) and out-of-plane (in which
case the energy is $\mu_0 \gamma_B M_S^2/2$).  Nano-elements have
additional side surfaces and it was recently proposed by Kohn and
Slastikov\cite{KS05} to use a similar expression for local surface
anisotropy of magnetostatic origin on all surfaces with some {\em a priori}
unknown constant $K_s$, replacing the $m_Z$ by a normal magnetization
component on the surface. Let us try this approach.

\subsection{Local anisotropy approximation}
When vortex is completely inside the particle and is centered ($a=0$) the meron
does not contribute to the anisotropy energy, and the contribution of soliton
part is
\begin{equation}
\label{eq:eain}
e_\mathrm{A}^{in}=\frac{1}{R^2}\int\limits_0^{R_V}
\frac{\left(r^2-R_V^2\right)^2}{\left(r^2+R_V^2\right)^2} r \ud r=
\frac{R_V^2}{R^2} \frac{\left( 3 - 4 \log 2 \right)}{2}.
\end{equation}
The total energy density $e_\mathrm{EX}+e_\mathrm{A}^{in}$ at $a=0$
now has a minimum when $R_V^A=L_E/\sqrt{3 - 4 \log 2}$. Or,
approximately, $R_V^A\approx 2.09698 L_E = 0.59155 \sqrt{4\pi} L_E$,
independent on cylinder's thickness $L_Z$. The thickness independence
is the result of expressing the magnetostatic energy in the form of
surface anisotropy, and, as will be seen later, is
wrong. Nevertheless, unlike the purely exchange approximation, the
vortex core size is now stable. It is also worth noting that local
anisotropy approximation is exact in the limit of vanishing film
thickness, so that
$R_V^A/L_E=\lim_{\lambda\rightarrow0}\rho_V(\lambda)$, where
$\rho_V(\lambda)$ is the vortex core radius, computed with full
treatment of magnetostatics (\ref{eq:eqrhov}).

But stable vortex core size is not all, the vortex must also be stable
with respect to the displacement of its
center. Magnetostatically-induced anisotropy on the cylinder's face
does not stabilize the vortex, since (for the case of vortex inside
the particle) its energy is independent on the vortex center
displacement $a$. But the exchange energy (\ref{eq:exchange_total})
decreases when vortex is displaced ($|a|$ increases from 0), which
leads to instability. To consider this case properly within the local
anisotropy approximation let us follow the proposal of Kohn and
Slastikov\cite{KS05} and introduce additional surface anisotropy $K_S$
on the cylinder's side. It gives the following contribution to the
energy of displaced vortex, assuming it is fully within the particle
\begin{equation}
E_\mathrm{A}^{s}=K_S L_Z R \!\int\limits_0^{2\pi}\!\left(\uRe\frac{e^{-\imath
\varphi} f(R e^{\imath \varphi})}{|f(R e^{\imath \varphi})|}\right)^2\!\!\!
\ud \varphi =\frac{K_S L_Z \pi a^2}{R},
\end{equation}
where $\uRe$ takes the real part of its right argument.  The exchange
energy (\ref{eq:exchange_meron_uniform}) can be expanded as
$E_\mathrm{EX}(a)\approx \text{\it const} - C L_Z \pi a^2/(2
R^2)$. Equating two second order terms in $a$ gives the condition for
vortex stability with respect to displacement: $R>R_S^A=C/(2
K_S)$. For radii, smaller than $R_S^A$, the vortex is unstable. This
is, again, only partially correct. The stability condition turns out
to be independent on $L_Z$, which means that, while the particles of
very small radii are correctly single-domain, in particles of
disappearing thickness the vortex state is unconditionally stable,
which is qualitatively wrong. Nevertheless, if one deals with
particles of specific size and considers $K$ and $K_S$ as free
parameters, the approach of Kohn and Slastikov\cite{KS05} may yield a
reasonable approximation to the stability and evolution of vortex
state, in this case $K_S$ will have to vanish as $L_Z \rightarrow
0$. The advantage of local anisotropy approximation is simplicity, as
it allows to get explicit expressions for most interesting
quantities. The full account for long-range magnetostatic interaction,
which is necessary to build the vortex state theory without free
parameters, is much harder to do. Yet, in the following text,
approximate expressions for vortex radius and energy with full account
of magnetostatics are derived, which are almost as simple.

\subsection{Full magnetostatic energy evaluation}
To compute the magnetostatic energy let us use the magnetic charges
formalism, introducing a magnetic charge density $-\mathrm{div} \,
{\vec m}$ which is automatically equal to the normal component of
magnetization on the surface of magnetic material (in which case it is
a surface charge density $\sigma$). In centered vortex $a=0$ there is
only a face charge (surface charge on cylinder's face, proportional to
$m_z$), equal to
\begin{equation}
 \sigma(r) = M_S \frac{R_V^2-r^2}{R_V^2+r^2},
\end{equation}
where $r<R_V$ because all the charge is concentrated in the vortex core. The
interaction energy of two such systems of charge at parallel planes (faces of
the cylinder), separated by distance $h$, can be directly written as
\begin{equation}
 \frac{4 \pi U(h)}{\mu_0 \gamma_B}=\!\!\int\limits_0^{R_V} \!
\int\limits_0^{2\pi} \! \int\limits_0^{R_V} \!
\int\limits_0^{2\pi} \!\!
\frac{\sigma(r_1)\sigma(r_2) r_1\ud r_1\ud\varphi_1 r_2\ud r_2\ud\varphi_2}
{\sqrt{r_1^2\!+\!r_2^2\!-\!2 r_1 r_2 \cos (\varphi_1\!-\!\varphi_2)\! +\! h^2}}.
\end{equation}
It is possible to obtain two equivalent representations for this integral,
one by directly integrating over the angles
\begin{eqnarray}
\int\limits_0^{2\pi}
\int\limits_0^{2\pi}
\frac{\ud\varphi_1\ud\varphi_2}
{\sqrt{r_1^2+r_2^2-2 r_1 r_2 \cos (\varphi_1-\varphi_2) + h^2}}= & &\nonumber \\
\frac{8 \pi K(\frac{4 r_1 r_2}{h^2+(r_1+r_2)^2})}
{\sqrt{h^2+(r_1+r_2)^2}} & &,
\end{eqnarray}
where $K(k)$ is a complete elliptic integral of the first kind, which gives
\begin{equation}
 \label{eq:uelliptic}
 u(\chi)=\frac{2}{\pi}
\int\limits_0^1 \int\limits_0^1 
\frac{\rho_1 (1-\rho_1^2) \rho_2 (1-\rho_2^2)  K(\frac{4 \rho_1
\rho_2}{\chi^2+(\rho_1+\rho_2)^2})}
{(1+\rho_1^2) (1+\rho_2^2)\sqrt{\chi^2+(\rho_1+\rho_2)^2}},
\end{equation}
where the dimensionless quantities $u=U/(\mu_0\gamma_B M_S^2\pi
R_V^3)$, $\chi=h/R_V$, $\rho_1=r_1/R_V$,
$\rho_2=r_2/R_V$ have been introduced.
Another representation can be obtained using summation theorem for Bessel's
functions of the first kind\cite{GM01}
\begin{equation}
\label{eq:ubessel}
 u(\chi)=
\int\limits_0^\infty e^{-k \chi}
\left(
\int\limits_0^1 \frac{1-\rho^2}{{1+\rho^2}} J_0(k \rho) \rho \ud \rho
\right)^2\ud k,
\end{equation}
which is shorter on paper, but, unlike (\ref{eq:uelliptic}), is, actually, a
triple integral. The magnetostatic energy of the vortex core is then
\begin{equation}
e_\mathrm{MS}=\frac{E_\mathrm{MS}}{\mu_0\gamma_B M_S^2\pi
L_Z R^2}=\frac{R_V^3}{L_Z R^2}(u(0)-u(L_Z/R_V)),
\end{equation}
where the first term accounts for the face charge's
self energy, while the second for interaction of charges on the opposite faces.
The total dimensionless energy of the cylinder with centered vortex is
\begin{equation}
  \label{eq:dimcent}
 e = \frac{L_E^2}{R^2} \left(2 - \log \frac{R_V}{R}\right)+
e_\mathrm{MS} .
\end{equation}
Minimization of this energy with respect to $R_V$ results in the following equation
\begin{equation}
  \label{eq:eqrhov}
 -\frac{1}{\rho_V}+\frac{3 \rho_V^2
(u(0)-u(\lambda/\rho_V))}{\lambda} + \rho_v u'(\lambda/\rho_V) = 0,
\end{equation}
where $\rho_V=R_V/L_E$, $\lambda=L_Z/L_E$ and prime means derivative. This
equation in independent on particle radius.

\section{Approximate expressions for vortex core radius and energy}
It is, of course, possible to evaluate the integrals (\ref{eq:uelliptic}),
(\ref{eq:ubessel}) on computer (but even this is tricky, since the second is a
badly converging oscillating improper integral and the first contains a
peak at $\rho_1=\rho_2$, turning into a line of
integrable logarithmic singularities when $h=0$) and solve the transcendental
equation (\ref{eq:eqrhov}) numerically, but it is far less convenient (and
useful), compared to having their simple analytical expressions. Let us now
obtain such expressions approximately.

The simplest is the case of large cylinder thickness, corresponding to
$\chi \gg 1$. In this case the outer integral in (\ref{eq:ubessel}) is
converging very fast, and also the integrand in (\ref{eq:uelliptic}) is well
behaved. This allows to perform straightforward Taylor's expansion of
the integrand and perform the integration term by term, which gives:
\begin{eqnarray}
u(\chi)& = & \frac{(\log (4)-1)^2}{4 \chi }+\frac{3+8 \log ^2(2)-10
\log (2)}{8 \chi^3}+\nonumber \\
  & & \frac{35+4 \log (2) (18 \log (2)-25)}{32 \chi
^5}+ \ldots
\end{eqnarray}
Solving (\ref{eq:eqrhov}) with this $u$ results in the following expansion
for the equilibrium vortex core radius
\begin{equation}
  \label{eq:rhoeqlargelambda}
\rho_V^\mathrm{EQ}(\lambda)= \left(\!\frac{\lambda}{3 u_0}\!\right)^{\!\!1/3}\!\!\!\!
%\frac{\lambda^{1/3}}{(3 u_0)^{1/3}}
+
\left(\!\frac{(\log (4)-1)^6}{81^{2}u_0^{5}\lambda}\!\right)^{\!\!1/3}\!\!\!\!
%\frac{(\log (4)-1)^2}{81^{2/3}u_0^{5/3}\lambda^{1/3}}
+
\frac{(\log (4)-1)^4}{81\lambda u_0^3}
+ \ldots,
\end{equation}
where $u_0=u(0)=0.0826762$. Substituting it into (\ref{eq:dimcent})
gives the equilibrium energy of thick cylinder ($\lambda \gg 1$) with
magnetic vortex
\begin{equation}
\label{eq:eeqlargelambda}
%\frac{E^\mathrm{EQ}}{\mu_0\gamma_B M_S^2 \pi L_Z R^2}
e^\mathrm{EQ}\approx
\frac{7+\log \left(\frac{3 \rho^3 u_0}{\lambda }\right)}{3 \rho
^2}-\frac{(\log (4)-1)^2}{12 \rho ^2 (3
u_0^{4} \lambda^{2})^{1/3}},
\end{equation}
where $\rho=R/L_E$.

These expressions are simple and for $\lambda>2$ are precise to a few
percent (and for $\lambda/\sqrt{4\pi}>1$ the precision is better than
1\%). The problem, however, is that assumption of uniformity of
magnetic texture along $Z$ axis is not a good approximation for thick
cylinders ($\lambda \gg 1$), which, eventually, start to develop a 3D
structure (such as variation of vortex radius with $Z$ at first). In
other words, the expressions
(\ref{eq:rhoeqlargelambda}),(\ref{eq:eeqlargelambda}) are precise
mostly in the region, where the vortex (\ref{eq:f}) can be far from
the ground state of the system (Eq. \ref{eq:eeqlargelambda} will still
be useful for finding the extent of this region by comparing it to the
energy of other magnetization textures).

It might be tempting to expand the magnetostatic function $u(\chi)$
around $\chi=0$ and build the approximate vortex state theory on top
of that. The difficulty is that $u(\chi)$ is not analytic at $\chi=0$
(it has terms, proportional to $\chi \log \chi$) and the integrals for
higher order Taylor expansion terms do not converge. Since the very
thin particles are single-domain and also thin cylinders with large
radius start to develop a domain structure or several vortices, bound
as finite fragments of cross-tie domain walls, such approximation
would also be the most precise in the region, where the vortex is not
the ground state.

This suggests the idea to build the magnetostatic function expansion
around an intermediate point $\chi=1$, where the function $u(\chi)$ is
analytic and otherwise well-behaved. Such expansion is most precise
for $R_V \sim L_Z$, where all the physical assumptions of the vortex
state theory are valid. This region is also close to the triple point
on the magnetic phase diagram\cite{MG02_JEMS}.  The point $\chi=1$
corresponds to
\begin{equation}
\lambda_0=\frac{1}{\sqrt{3(u(0)\!-\!u(1))\!+\!u'(1)}}\approx2.7284 \approx
0.76967\sqrt{4 \pi} .
\end{equation}
The expansions for equilibrium vortex radius and energy about the point
$\lambda_0$ are the following
\begin{eqnarray}
\rho_V^\mathrm{EQ}(\lambda)& = &\sum\limits_{i=0}^\infty a_i
(\lambda-\lambda_0)^i \\
%\frac{E^\mathrm{EQ}}{\mu_0\gamma_B M_S^2 \pi L_Z R^2} 
e^\mathrm{EQ}
& = &
\frac{\log (\rho/\lambda_0)}{\rho^2}+
\frac{1}{\rho^2} \sum\limits_{i=0}^\infty b_i
(\lambda-\lambda_0)^i,
\end{eqnarray}
where the first few coefficients are
\begin{center}
\begin{tabular}{l | l | l | l | l | l }
\hline
$i$   & 0          & 1         & 2        & 3       & 4       \\ \hline
$a_i$ & $\lambda_0$& 0.189400 &-0.012521& 0.001182&-0.000093 \\ \hline
$b_i$ & 2.387556   &-0.082425 & 0.010332&-0.00171 & 0.000315 \\
\hline
\end{tabular}
\end{center}
While it is possible to write down analytical expressions for these
coefficients, they are highly unwieldy and add little value. In the
above numerical form these coefficients are just universal
dimensionless constants, they decay rapidly and are sufficient to
compute the equilibrium vortex core radius and energy with precision
up to $0.5\%$ in the interval $0<\lambda<6$. Comparison of the above
analytical approximations with exact numerical values of core radius
and energy are shown in Figure.
\begin{figure}
\label{fig:energyradius}
 \centering{\includegraphics[scale=0.5]{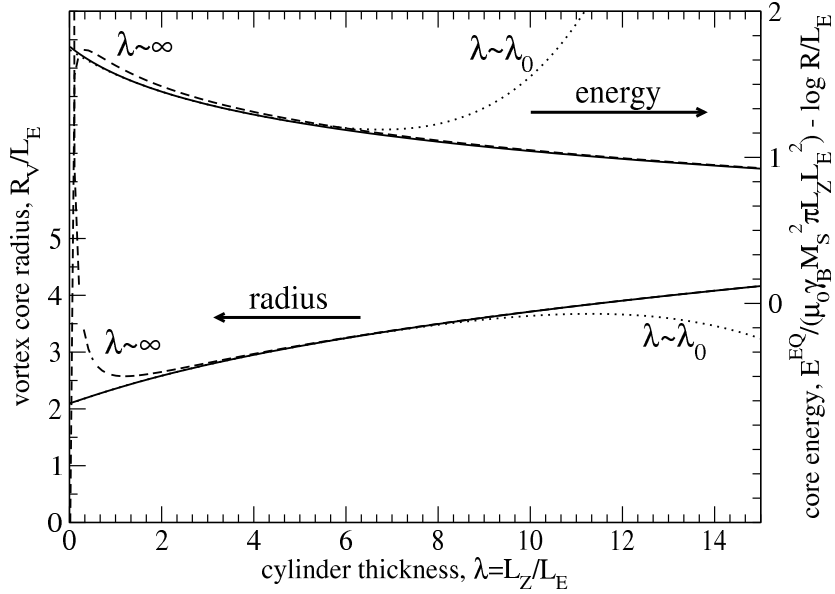}}
\caption{Vortex core radius and energy density, as function of the cylinder
thickness. The dotted and dashed lines show approximate analytical expressions,
the solid lines are the result of the exact numerical calculation.}
\end{figure}

\section*{Conclusions}
Starting with the description of magnetization distributions via
complex variable, various simplifying physical approximations for
magnetic Hamiltonian were consistently (in the same notations and
units) presented and compared with their advantages and deficiencies
highlighted. Formulas for the exchange energy of such distributions
(\ref{eq:exchange_soliton_final})-(\ref{eq:exchange_meron_general}),
provided the vortices and anti-vortices are fully contained inside the
particle (which covers many distributions of Ref.~\cite{M10}),
were presented here for the first time. Two simple and explicit
analytical approximations for equilibrium vortex core radius and
energy in circular cylinder were derived, which, together, cover the
whole range of cylinder geometries.

%\bibliographystyle{elsarticle-num}
%\bibliography{klm_base}

\end{document}